\documentclass[aps,pre,amsmath,amssymb,twocolumn,showpacs,superscriptaddress]{revtex4}     
\usepackage{graphicx}

\begin{document}

\title{Fractional Stokes-Einstein and Debye-Stokes-Einstein
  relations in a network forming liquid}

\author{Stephen R. Becker} \thanks{Present Address: Department of
  Applied and Computational Mathematics, California Institute of
  Technology, Pasadena, CA 91125} \affiliation{Department of Physics,
  Wesleyan University, Middletown, CT 06459, USA} 

\author{Peter H. Poole} \affiliation{Department of Physics,
  St. Francis Xavier University, Antigonish, Nova Scotia B2G 2W5,
  Canada} 

\author{Francis W. Starr} \affiliation{Department of Physics, Wesleyan
  University, Middletown, CT 06459, USA}

\date{May 7, 2006}

\begin{abstract}
  We study the breakdown of the Stokes-Einstein (SE) and
  Debye-Stokes-Einstein (DSE) relations for translational and
  rotational motion in a prototypical model of a network-forming
  liquid, the ST2 model of water.  We find that the emergence of
  ``fractional'' SE and DSE relations at low temperature is ubiquitous
  in this system, with exponents that vary little over a range of
  distinct physical regimes.  We also show that the same fractional SE
  relation is obeyed by both mobile and immobile dynamical
  heterogeneities of the liquid.
\end{abstract}

\maketitle

The translational ($D$) and rotational ($D_{r}$) diffusion constants
of a macroscopic object in a simple liquid are well-characterized by
the Stokes-Einstein (SE) relation, $D\tau/T=c$, and the
Debye-Stokes-Einstein (DSE) relation, $D_{r}\tau/T=c_r$.
In these relations, $\tau$ is a relaxation time proportional to the
viscosity of the liquid, $T$ is the temperature, and the value of the
constants $c$ and $c_r$ depend on the geometry of the object and the
boundary conditions.  While these relations were originally formulated
for the diffusion of macroscopic objects, they are known to hold when
the diffusing object is itself a molecule of the
liquid~\cite{egelstaff,SE-obedience}.  Hence the SE and DSE relations
provide a simple connection between mass and momentum transport in a
liquid.

As $T$ decreases and the liquid becomes increasingly viscous (assuming
crystallization is avoided), numerous experiments have shown a failure
of the SE relation for $T \lesssim 1.3 T_{g}$, where $T_{g}$ is the
glass transition
temperature~\cite{pollack,ediger-rev,sillescu-rev,fujara,ediger,sillescu}.
Specifically the ratio $D\tau/T$ is found to increase by as much as 2
or 3 orders of magnitude on cooling~\cite{sillescu}.  Translational
diffusion is thus said to be enhanced relative to viscosity.  When the
SE relation fails, it has been empirically found that a ``fractional''
Stokes-Einstein (F-SE) relation, $D \sim (\tau/T)^{-\xi}$
holds for a wide range of
liquids~\cite{pollack,ediger,bocquet,ehlich,voronel}, where $0.5 \le
\xi \le 0.95$, with values between $0.7$ and $0.8$ being most commonly
reported.  While a definitive theoretical prediction of the value of
$\xi$ has been elusive, various theoretical models predict $\xi$ in
the same range~\cite{jack,coniglio,schweizer,cg1}.  The situation is
more complex for the DSE equation.  In the same $T$ range where the SE
equation fails, the DSE equation has been experimentally found to be
valid for most liquids~\cite{ediger-rev,sillescu-rev}.  However, in
other experiments~\cite{sillescu,dino-exp,poles} a fractional
Debye-Stokes-Einstein (F-DSE) relation, $D_{r} \sim
(\tau/T)^{-\xi_r}$, has been observed.  Further complicating the
situation is the fact that $D_{r}$ is experimentally inaccessible, and
experiments must test the DSE relation using a re-orientational
correlation time, which may not be simply related to $D_{r}$.
Molecular dynamics simulations can directly evaluate $D_{r}$, and
simulations of dumbbell molecules show a failure of the DSE relation
using $D_{r}$~\cite{dino}.

A commonly proposed explanation for the breakdown of the SE relation
is the presence of dynamical heterogeneity
(DH)~\cite{stillinger,tarjus}, i.e. spatially correlated regions of
relatively high or low mobility that persist for a finite lifetime in
the liquid, and that grow in size as $T$ decreases.  The existence of
DH has been confirmed and quantified in both
experiments~\cite{ediger-rev,sillescu-rev} and
simulations~\cite{glotzer-rev}.  However, only a few liquid simulation
studies directly probe the relationship between DH and the breakdown
of the SE relation~\cite{sanat,berthier}.

\begin{figure}
\centerline {\includegraphics[width=3.4in]{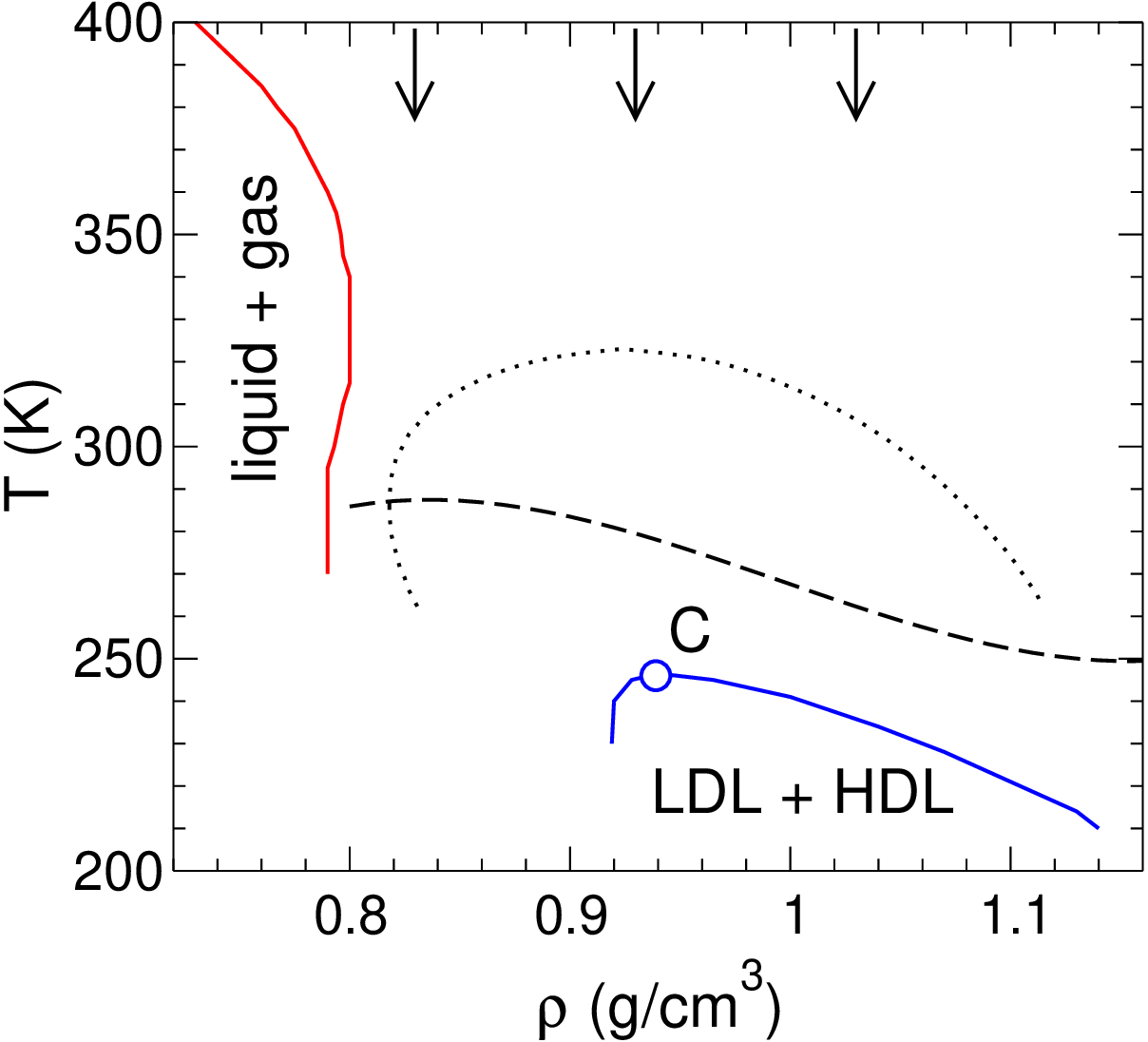}}
\caption{ ST2 phase behavior in the $\rho$-$T$ plane.  Solid curves
  show spinodals bounding regions of liquid-gas and HDL-LDL liquid
  phase separation.  $C$ is the critical point of the HDL-LDL
  transition.  The three arrows indicate the densities focused on in
  subsequent plots. The dotted line is the locus of density extrema.
  The dashed curve is the locus of points at which $D\tau/T$ is 1.5
  times its value at $400$~K, along isochores.}
\label{fig:PD}
\end{figure}

In this Letter, we study the breakdown of the SE and DSE relations in
simulations of the ST2 model of water~\cite{st2}.  The ST2 model is a
tetrahedral arrangement of charges, centered within a Lennard-Jones
envelope, that qualitatively reproduces a number of water's anomalies.
However, our goal is not to elucidate the properties of water
specifically; indeed there are many water models more accurate than
ST2.  Rather, we choose this model because it is known to display a
diversity of extreme liquid behavior within a single system
(Fig.~\ref{fig:PD}).  At high density $\rho$, ST2 behaves as a typical
fragile glass-forming molecular liquid.  However, near
$\rho=0.93$~g/cm$^3$ and $T=240$~K, a liquid-liquid critical point
occurs in the equation of state, demarcating the onset of phase
separation between a low density liquid (LDL) and a high density
liquid (HDL) phase~\cite{PSES92}.  At still lower $\rho$, near
$0.83$~gm/cm$^3$, the behavior of ST2 is dominated by the emergence of
an increasingly well-structured random tetrahedral network (RTN) as
$T$ decreases.  The emergence of the RTN is associated with the onset
of a fragile-to-strong crossover in the liquid transport
properties~\cite{GP}.  We study each of these three distinct regimes,
to test how the character of SE and DSE breakdown changes as the
behavior of the liquid state itself changes. We also test if the
origins of the F-SE relation can be found in the DH of the system.

\begin{figure}
\centerline {\includegraphics[width=3.4in]{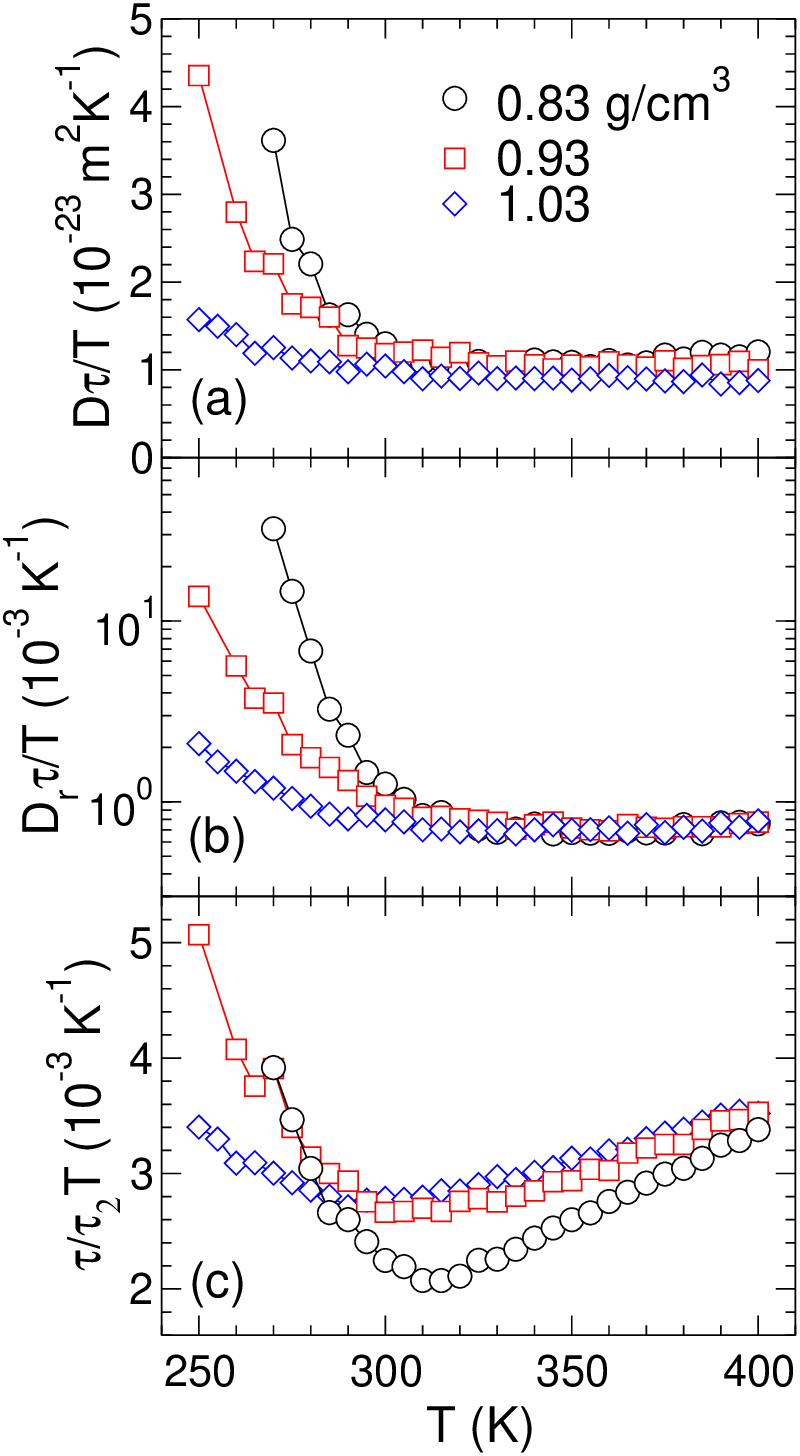}}
\caption{(a) Breakdown of the SE relation at low T.  (b) Test of the
  DSE relation using $D_r$.  (c) Test of the DSE relation using
  $1/\tau_2$ as a proxy for $D_r$.}
\label{fig:SE}
\end{figure}

\begin{figure}
\centerline {\includegraphics[width=3.4in]{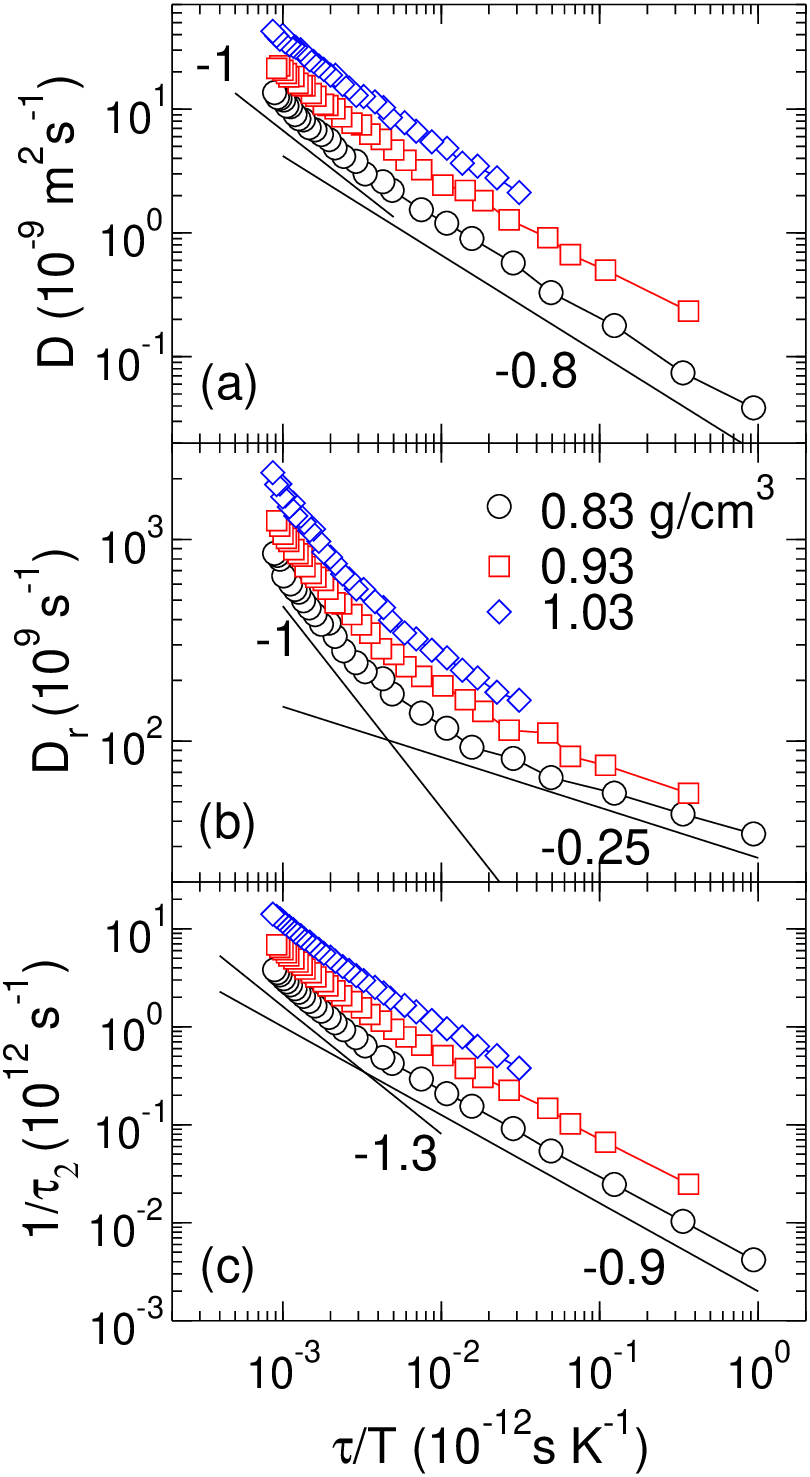}}
\caption{Testing for (a) F-SE behavior; (b) F-DSE behavior using
  $D_r$; and (c) F-DSE behavior using $1/\tau_2$.  Curves for
  $\rho=0.93$ and $1.03$~g/cm$^3$ have been multiplied by arbitrary
  factors to facilitate comparison. Solid lines have the slopes
  indicated. }
\label{fig:FSE}
\end{figure}

\begin{figure}
\centerline {\includegraphics[width=3.4in]{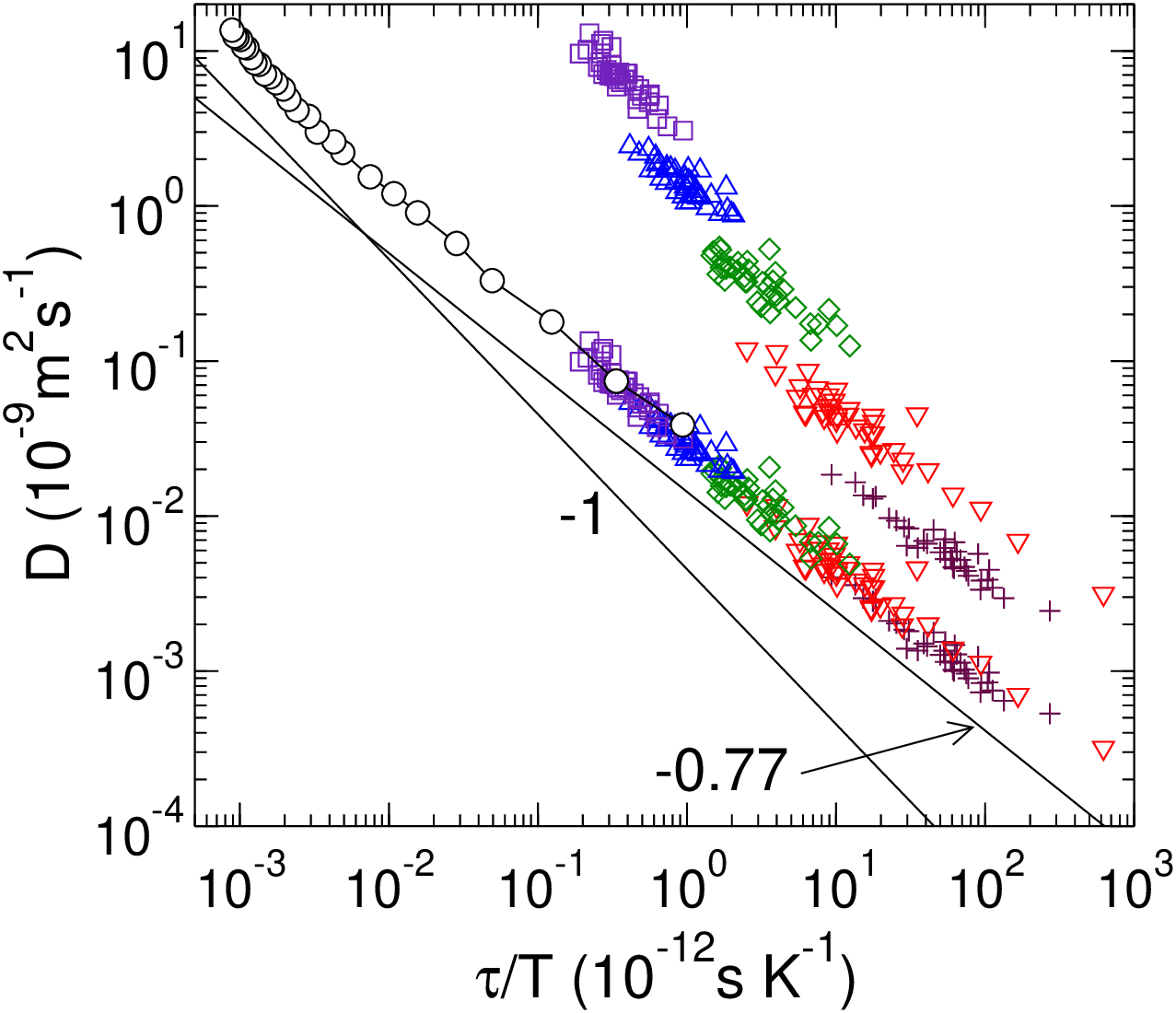}}
\caption{Testing for F-SE behavior at $\rho=0.83$~g/cm$^3$. Circles
  are data from $N=1728$ simulations.  The ``clouds'' of other symbols
  are obtained from ensemble simulations of a $N=216$ system at
  255~K~($+$), 260~K~($\bigtriangledown$), 265~K~($\diamond$),
  270~K~($\bigtriangleup$) and 275~K~($\Box$).  The ``cloud'' for each
  $T$ has been repeated and shifted by an arbitrary factor to reveal
  its shape in isolation from the other clouds. Solid lines have the
  slopes indicated.}
\label{fig:DH}
\end{figure}


Unless otherwise indicated, our data are generated by molecular
dynamic simulations using $N=1728$ molecules with a time step of 1~fs.
Simulations consist of equilibration, followed by a production phase,
each of which is run for the longer of 100~ps or the time needed for
the mean-squared displacement per molecule $\langle r^{2} \rangle$ to
reach 1~nm$^2$ -- roughly three molecular diameters.  Simulations are
carried out in the range $250$~K$\le T \le 400$~K at intervals of 5~K
and for $0.80$~g/cm$^3\le \rho \le 1.20$~g/cm$^3$ at intervals of
$0.01$~g/cm$^{3}$.  For simplicity of the figures, we only present
data along three representative isochores: (i) $\rho
=0.83$~g/cm$^{3}$, where the RTN structure dominates; (ii) $\rho
=0.93$~g/cm$^{3}$, approximately the critical density for the
liquid-liquid transition; and (iii) $\rho =1.03$~g/cm$^{3}$ at which
the system behaves more as a simple fragile liquid.  The simulations
are carried out at fixed $\rho$ and $T$, employing the Berendsen heat
bath with a time constant of 2~ps.  Electrostatic interactions are
truncated at 0.78~nm and the energy and pressure are corrected using
the reaction field method~\cite{allen-tildesley}.

We evaluate $D$ from the asymptotic behavior of $\langle r^{2} \rangle
= 6 Dt$; and $D_{r}$ by tracking the mean-squared angular displacement
$\langle \phi^{2} \rangle$ of the molecular dipole moment vector and
use the asymptotic relation $\langle \phi^{2} \rangle = 4
D_{r}t$~\cite{marco}.  The alpha-relaxation time $\tau$ is defined
as the time where the normalized coherent intermediate-scattering
function decays to a value of $1/e$, evaluated at the closest discrete
wave number to $q = 18$~nm$^{-1}$, the approximate location of the first peak in
the static structure factor.  The dipole relaxation time $\tau_{2}$ is
defined as the time at which the correlation function
$\langle P_{2}[\cos\theta(t)]\rangle$ decays to a value of $1/e$, where $P_{2}$ is
the second Lengendre polynomial and $\theta(t)$ is the angle defined
by the orientations of the dipole moment vector of a molecule at times
$t$ and zero.

To determine in which regions of the phase diagram the SE and DSE
relations are valid, we show the behavior of $D\tau/T$ and $D_r\tau/T$
in Fig.~\ref{fig:SE}.  For $T \gtrsim 300$~K, Figs.~\ref{fig:SE}(a)
and (b) show that both the SE and DSE ratios are nearly constant, as
expected.  The growth of the SE relation reflects the expected
enhanced translational diffusion.
In the same sense, we also find enhanced rotational diffusion; noting
that Fig.~\ref{fig:SE}(b) employs a log scale, the enhancement of
rotational motion exceeds that of translational motion.  This is
consistent with spin-lattice relaxation experiments on
water~\cite{ludemann}.  The $T$ at which the SE relation breaks down
is weakly $\rho$ dependent, as shown in Fig.~\ref{fig:PD}.

The deviation from the DSE relation can be seen as somewhat surprising
since many (but not all) experiments have found that the DSE relation
appears to be valid even when the SE relationship
fails~\cite{ediger-rev,sillescu-rev}.  In experimental studies,
$D_{r}$ cannot be directly measured. As a result, the inverse dipole
relaxation time $\tau_{2}^{-1}$ is frequently used as a substitute,
since it can be readily measured.  Since we found a deviation from the
DSE relation using $D_{r}$, we consider whether the using $\tau_{2}$
yields different results, as shown in Fig.~\ref{fig:SE}(c).
Curiously, using $\tau_{2}$ we find no $T$ range where the DSE
relation appears valid.  
For the system we study, the difference found using $\tau_2$ versus
$D_r$ may result from large reorientational motions associated with
hydrogen bond switching. Whatever the cause, Fig.~\ref{fig:SE}(c)
illustrates that substitution of $D_{r}$ with $\tau_{2}$ may not, in
general, be appropriate~\cite{dino-exp}.

To test if the SE relation is replaced by the F-SE form at low $T$, we
parametrically plot the relation between between $D$ and $\tau/T$ in
Fig.~\ref{fig:FSE}(a).  We find that there are two distinct regions:
(i) $T\gtrsim 275$ where we find an exponent $\xi = 1$, consistent
with the SE relation, and (ii) lower $T$ where we find a F-SE relation
with an exponent in the commonly observed range $0.7 \le \xi \le 0.8$.
Moreover, the value of $\xi$ is nearly independent of $\rho$.  Since
the DSE relation is also clearly violated in this system, we check for
a F-DSE relation in Fig.~\ref{fig:FSE}(b).  Indeed, analogous to
Fig.~\ref{fig:FSE}(a), we find that a high-$T$ DSE regime is replaced
at low $T$ by a F-DSE relation with $\xi_r \approx 0.25$,
significantly smaller than $\xi$ for the F-SE relation.  The smaller
value of $\xi_r$ compared to $\xi$ is also observed in experiments on
OTP~\cite{dino-exp} and is consistent with our finding that the
rotational motion is more dramatically enhanced than the translational
diffusion.  Like $\xi$, we find that $\xi_r$ varies weakly with
$\rho$, if at all.  Hence, despite the wide variation of liquid state
properties over the density range studied here, the quantitative form
of the F-SE and F-DSE behaviors is strikingly uniform.

We next analyze the behavior of the ST2 model to test for a connection
between DH and F-SE behavior.  We focus on the $\rho =
0.83$~g/cm$^{3}$ isochore, and conduct $40$ independent simulations
with $N=216$ molecules at each $T$ for $255\le T \le 275$~K, at $5$~K
intervals, again using the criterion that each run continues until
$\langle r^{2} \rangle = 1$~nm$^2$.  The smaller system (compared
to $N=1728$) allows us to probe much longer time scales and to
generate an ensemble of many runs in a computationally accessible
time.  As DH emerges at low $T$, each member of the ensemble at the
same $T$ becomes more strongly influenced by transient mobile and
immobile regions.  As a consequence, some members of the ensemble take
a longer time to reach the run-time criterion, and some shorter.
Thus, we can study the differences between systems dominated by mobile
and immobile DH subdomains without the need to classify individual
molecules by their mobility.

It has been hypothesized that the SE violation is due primarily to the
presence of transient mobile
regions~\cite{stillinger,tarjus,ediger-rev}.  In this picture, a
``background'' of largely immobile regions obeys the normal SE
behavior, while mobile regions violate the SE relation.  If true, we
should find that the ensembles members that are dominated by mobile
regions should exhibit a prominent F-SE behavior, while the members
dominated by immobile regions should tend to follow a normal SE
behavior.  In this case, a parametric scatter plot of $D$ against
$\tau/T$ (Fig.~\ref{fig:DH}) for all ensembles should show a local
curvature of the data for each ensemble at a given $T$, such that the
low-$D$ envelope of the data reflects SE behavior, and the high-$D$
envelope gives F-SE behavior.  This is not what Fig.~\ref{fig:DH}
shows. Rather, we find that all systems, both those that are
relatively fast and those that are slow, all obey the same F-SE
relation with $\xi \approx 0.77$.  Therefore, for the ST2 model of
water at $\rho=0.83$~g/cm$^3$, both the mobile and immobile regions of
the liquid deviate in the same way from the SE relation.  In this
sense, the violation of the SE relation is spatially homogeneous,
despite the existence of DH in the system.  

We note that $\xi = 0.77$ matches the prediction of the entropic
barrier hopping theory of Ref.~\cite{schweizer}, although
Ref.~\cite{schweizer} assumes the SE relation is valid over small
domains, and that FSE behavior arises from averaging over spatial
variation of the domains.  However, more recent work by the same
authors~\cite{schweizer2} observes a spatially ubiquitous FSE, as
found in this work.Very recently, Ref.~\cite{sanat} studied the impact
of heterogeneous dynamics on the SE relation in the hard sphere system
by identifying a subset of ``hopping''mobile molecules; their results
indicated that only the hopping molecules violate the SE relation,
though they did not investigate F-SE behavior.  It would be useful to
conduct a similar analysis of our system, in which hopping can also be
expected at low $T$.


Overall, our results demonstrate the remarkable robustness of the F-SE
and F-DSE behaviors to variations in both the mean and local molecular
environment.  The fractional behavior is observed across three
distinct physical regimes (fragile, critical, and network-forming),
with little variation of the exponents $\xi$ and $\xi_r$, indicating
an almost complete insensitivity to changes in the average liquid
structure.  Even the large static fluctuations associated with the
approach to a second order critical point do not observably effect the
manifestation of the F-SE and F-DSE relations.  We also find no
sensitivity of the F-SE behavior on the relative proportion of mobile
and immobile dynamical heterogeneities within a given system.  Both
locally fast and slow regions of the system obey the same F-SE
relation, so the phenomenon cannot be attributed to any one extreme
subset.  We emphasize that our results do not imply that F-SE behavior
and DH are disconnected, since the two phenomenon emerge in the same
range of $T$~\cite{bordat}, only that the relationship is not the anticipated one.
Our analysis also does not indicate a specific alternative origin for
F-SE and F-DSE behavior. However, it does suggest that in a successful
theory of the phenomena, F-SE and F-DSE behavior should emerge
as an intrinsic behavior of the liquid state as $T\to T_g$,
insensitive to the details of the structural and dynamical
fluctuations occurring in the system.  This is certainly consistent
with the nearly ubiquitous observation of SE breakdown in glass
forming liquids.

We thank J.\ Douglas, S.\ Kumar, D.\ Leporini, and K.\ Schweizer for
helpful discussion.  SRB and FWS thank the NSF for support under grant
number DMR-0427239.  PHP acknowledges the support of NSERC, CFI, AIF
and the CRC Program.  We thank G. Lukeman and the StFX hpcLAB for
computing resources and support.

\end{document}